\documentclass[12pt,twoside]{article}
\usepackage{fleqn,espcrc1,epsfig}
\newcommand{\bea}{\begin{eqnarray}}
\newcommand{\eea}{\end{eqnarray}}
\newcommand{\be}{\begin{equation}}
\newcommand{\ee}{\end{equation}}
\title{\vspace{-1.7cm}Charmonium Suppression -- Interplay of Hadronic and 
Partonic Degrees of Freedom
\thanks{This work was funded in part by the GSI and by the
Josef Buchmann Foundation.}}
\author{L.Gerland\address{Institut f\"ur Theoretische Physik,
 D-60054 Frankfurt am Main, Germany},
        L. Frankfurt${}^{\rm{a,}}$\address{School of Physics and Astronomy, Tel Aviv University,
69978 Ramat Aviv, Tel Aviv, Israel},
        M. Strikman\address{Department of Physics, Pennsylvania State University,
University Park, PA 16802, USA\\
and Deutsches Elektronen Synchrotron, DESY, Germany},
        H. St\"ocker${}^{\rm{a}}$,
        W. Greiner${}^{\rm{a}}$}
\date{\today}
\begin{document}
\maketitle
\begin{abstract}
Last year the E866-group of the Fermilab measured the
$x_F$ dependence of $J/\Psi$ and $\Psi'$ suppression in $pA$
collisions. We discuss two of the effects found in that
experiment with regard to color coherence effects: the 
different suppression of the $J/\Psi$ and the $\Psi'$
at $x_F<0$ and the significant suppression of both at large $x_F$.
The small $x_F$ regions is dominated by fully formed
charmonium states and thus enables us to discuss the
formation time and the cross section of the different
charmonium states. In the large $x_F$ region the interaction
of the charmonium states with nuclear matter has to be described
by partonic degrees of freedom, because 
in that kinematic domain the formation time is 
much larger than the nuclear radii. The understanding
of this region will be 
crucial for the interpretation
of the data of the future heavy ion colliders RHIC and LHC.  
\end{abstract}
\section{Introduction}
From considerations on formation times 
the interaction of charmonium 
states with a nucleus in a $pA$
collision can be roughly divided in three different kinematical regions.
The formation time is given by~\cite{brodsky}
\be
t_f(J/\Psi)={1\over E_{\Psi'}-E_{J/\Psi}}\approx 
{2 p \over M^2_{\Psi'}-M^2_{J/\Psi}}\approx
0.3\; {\rm fm/c}\cdot \gamma
\label{form}
\ee  
Here, $E_{\Psi',J/\Psi}$ ($M_{\Psi',J/\Psi}$) are the energies (masses) of the
$\Psi'$ and the $J/\Psi$ and $\gamma$ is the Lorentz-factor.
The formation time of 
a $\Psi'$ should be larger than that 
of a $J/\Psi$
by the ratio of the radii of these states, which is $\approx 2$.
In a recent letter~\cite{khar}
the attempt to extract the formation time from the experimental data
on $e^+e^- \rightarrow Q\bar Q$ has been made. The authors found 
$t_f(J/\Psi)=0.44\; {\rm fm/c}\cdot \gamma,\; t_f(\Psi')=0.91\; {\rm fm/c}\cdot \gamma$
which is slightly larger than the values used here.
Using eq.~(\ref{form}) in the rest frame of the nucleus 
leads to the following three kinematical regions:
\begin{itemize}
\item
hadronic: $t_f<1.8$ fm/c\\
The formation time is smaller than the average inter-nucleon distance in the
nucleus. Only fully formed charmonium states interact with the nucleons. In $pA$
collisions at $E_{lab}=800$ GeV this means $x_F<-0.3$.
\item
mixed: $1.8\; {\rm fm/c}<t_f<2\cdot R_A$\\
The formation time is larger than the average distance of the nucleons in the
nuclei but smaller than the diameter of the nucleus. The charmonium states 
already  can interact within their formation time but some of them are still formed
in the nucleus. In $pA$ ($E_{lab}=800$ GeV) this means $-0.3<x_f<0$ (the upper limit
depends of course on $A$, $x_f<0$ is for Be).
\item
partonic: $2\cdot R_A<t_f$\\
All charmonium states are formed beyond the nucleus. 
Hence the interaction has to be treated on the parton level.
\end{itemize}
In this work we discuss the hadronic and the mixed region. In section~\ref{model}
we describe the model and present the results.
A brief summary is given in section~\ref{sum}.

\section{Model and Results}
\label{model}

In~\cite{ger} we showed that the cross section 
of a fully formed $J/\Psi$ as predicted by pQCD~\cite{rady}
is much smaller than the cross section 
extracted from photoproduction processes
at $E_{lab}=20$ GeV~\cite{slac}
(here we have $\gamma=6$ and thus are in the hadronic region). We concluded that
the cross section of charmonium states at Fermilab energies is dominated by a 
non-perturbative contribution. 
At the same time the pQCD contribution to the cross section increases 
rapidly with energy. It will be relevant at RHIC and may even dominate at LHC.

To calculate the non-perturbative contribution to the charmonium
nucleon cross section (shown in Tab.~\ref{tab}) we used the 
parametrization $\sigma(b)=C\cdot b^2$.
Here, $b$ is the transverse diameter of the charmonium state (transverse
to the beam direction) and $C$ is a constant. 
This model is capable to describe the total cross sections of
$\pi - N,\;  K - N,\; \phi - N$ and $pp$ collisions.
We adjusted the constant $C={\sigma(\pi) / b^2(\pi)}$,
i.e. $\sigma(\pi)=25$ mb and
$b^2(\pi)=0.96$ fm${}^2$ to the cross section and the size of the pion.
(Note that the size of the pion is estimated
from form factor measurements
and thus is based on the idea of a non-relativistic continous charge density.
Thus it may appear dangerous 
to use it as a spatial distribution of current quarks.)

The transverse sizes of the charmonium states has been 
calculated with non-relativistic wave functions taken from \cite{eich,buch}. 
The results of this calculations differ by $\approx 25\%$. 
This difference illustrate the dependence of the prediction  
on the charmonium model. 
The subscripts of $\chi_{lm}$ are the quantum numbers of the spherical harmonics,
the angular momentum dependent part of the wave functions.
The $\chi$ mesons have two different cross sections because the transverse
size of P-states ($l=1$) depends on the third component of the angular momentum 
$m$: ${\sigma(\chi_{11})/ \sigma(\chi_{10})}=2$. This angular momentum dependent 
absorption leads to a polarization of the $\chi$-meson in $pA$ and $AB$ collisions.
\begin{table}
\centerline{\parbox[t]{4.5cm}{\begin{tabular}{|c|c|c|c|c|}
\hline
meson & $J/\Psi$ & $\Psi'$ & $\chi_{10}$ & $\chi_{11}$\\
\hline
(1) $\sigma$ [mb] & 2.5 & 10 & 3.8 & 7.6 \\
\hline
(2) $\sigma$ [mb] & 3.1 & 12.6 & 4.6 & 9.3\\
\hline
\end{tabular}}\hfill
\parbox[t]{8.8cm}{\vspace{-1cm}\caption{\small}
\vspace{-.3cm}
The non-perturbative charmonium nucleon cross section
(1) with the transverse size $b$ calculated with the wave function from
\cite{eich} and (2) from~\cite{buch}.
\label{tab}}}
\vspace{-1.1cm} 
\end{table}  

We assume in this work that the production of $c\bar c$-pairs is a hard
process, i.e.
$\sigma(pA\rightarrow c\bar c)= A\cdot \sigma(pp\rightarrow c\bar c)$. 
In the kinematic region discussed here, this assumption is
consistent with the data (for a recent review 
see e.g.~\cite{peng}) but may be inappropriate
at large $x_F$~\cite{duffy}.
Additionally we assume that the suppression of charmonium states is only
due to their final state interactions. In particular, we neglect 
the nuclear modifications of the parton distribution functions.

In the following we calculate the survival probability $S$ of a 
state $X$ in $pA$-collisions defined as $S_A^X=
{\sigma(pA\rightarrow X)}/({A \cdot \sigma(pN\rightarrow X)})$. Thus
$S=1$ for the production of $c\bar c$-pairs (the total charm cross section), 
with the assumptions made above. Our assumptions
allow to calculate the survival probability within the semiclassical
approximation of the Gribov-Glauber model as in~\cite{huf}:
\be
S_A^X={1 \over A} \int {\rm d}^2B\,{\rm d}z\, \rho(B,z)\cdot \exp \left(-
\int_z^{\infty}\sigma_{X} \rho(B,z'){\rm d}z'\right)\; ,
\ee
where $B$ is the impact parameter, $z$ the beam direction,
$\rho$ the density of the nucleus and $\sigma_X$
the cross section for a $XN$ collision. 

Now, we can calculate $S$ in the kinematical region that
we called hadronic in the introduction. However, we also want to discuss
the mixed region and therefore we have to describe the $XN$ cross section
within the formation time of the state $X$. We assume that the state $X$
is produced as a small color singlet $c\bar c$ pair that expands during
its formation time.
Interesting physics related to the propagation of color octet $Q\bar Q$
pairs is beyond our scope since it
requires methods beyond the semiclassical approximation.
For the region of negative $x_F$ the propagation of color octets should
be suppressed because 
its energy is insufficient to separate color from the quark-gluon
environment.
The cross section then reads~\cite{farrar}:
\bea
\sigma(z,z')_X & = & \sigma(z)+{z'-z \over t_f}(\sigma_X-\sigma(z))
\;{\rm for}\;z'-z<t_f\;\cr
\sigma(z,z')_X & = &\sigma_X 
\hspace{4.1cm}
{\rm otherwise}.
\label{expand}
\eea
Here, $\sigma(z,z')_X$ is the cross section of the state $X$ produced at the position $z$
after it moved to $z'$;
$\sigma(z)=1$ mb is the cross section at the production point $z$ and
$\sigma_X$ is the cross section of the state $X$ after the formation time
elapsed. That means $\sigma(z,z')_X\propto t$ and thus $b\propto \sqrt{t}$
as it should be for the expansion of wave packages in 
nonrelativistic quantum mechanics~\cite{brodsky}. 
This behaviour is called quantum diffusion.

\begin{figure}[t] 
\vspace{-1cm} 
\centerline{\parbox[b]{7cm}{\epsfxsize=9.cm 
\epsfbox{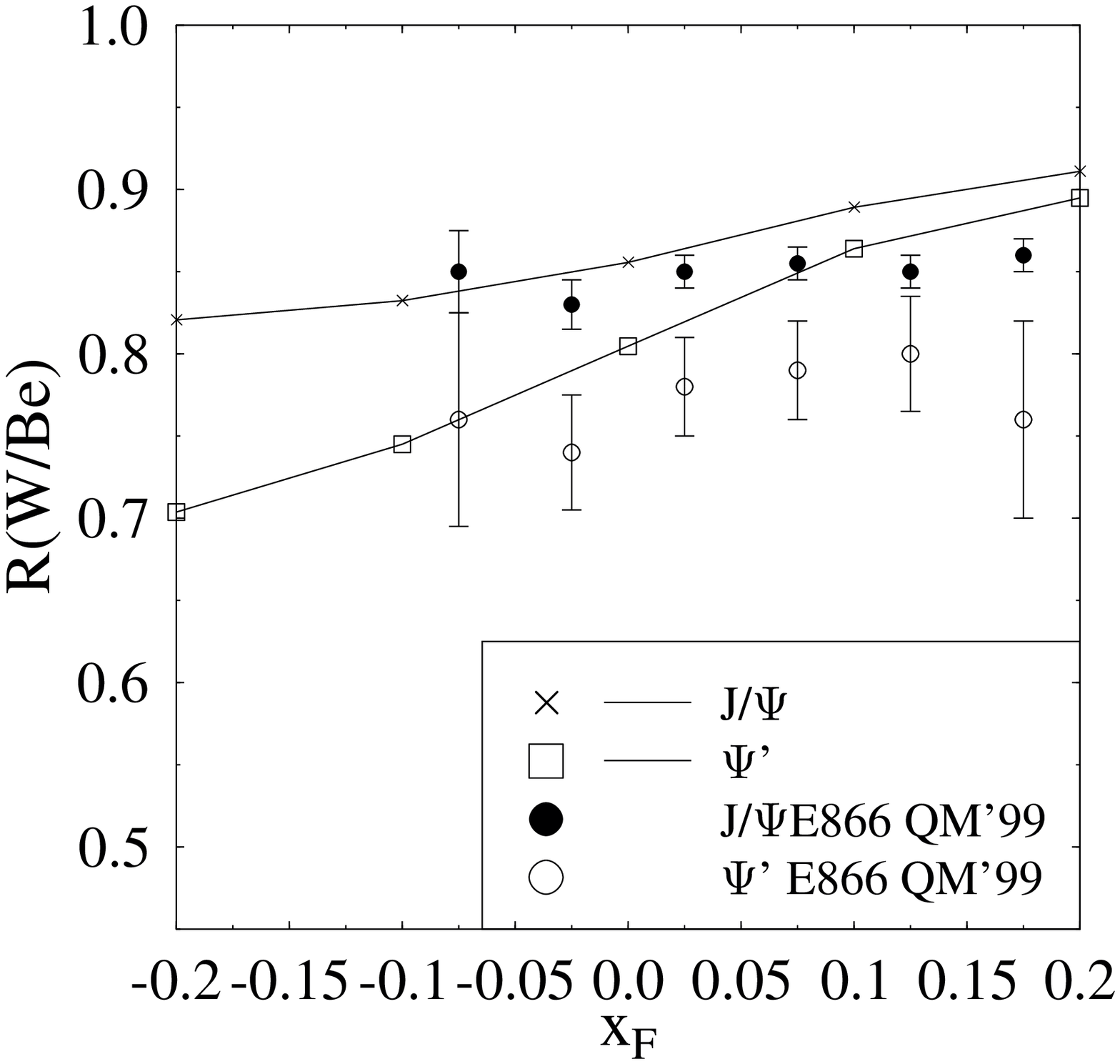}}
\hfill 
\parbox[b]{6.3cm}{\caption{\small} The ratios $R(W/Be):={S(pW)/ S(pBe)}$
of the $J/\Psi$ and the $\Psi'$ are plotted versus $x_F$ at $E_{lab}=800$
GeV. The used model (see text) is only valid for $x_F<0$. In this region the
calculation is in good agreement with the data.
\vspace{3cm}\label{ratio}}} 
\vspace{-1.8cm} 
\end{figure}

In Fig.~\ref{ratio} $R(W/Be):={S(pW)/ S(pBe)}$ is plotted versus $x_F$
for the $J/\Psi$ and the $\Psi'$. Note that a large fraction of the 
$J/\Psi$'s comes from the decay of $\Psi'$ and $\chi$, therefore:
\be
S=0.6\cdot ( 0.92\cdot S^{J/\Psi}+0.08\cdot S^{\Psi'})+0.4\cdot S^{\chi}\; ,
\ee
where $S^{J/\Psi}$ is the genuine survival propability of the $J/\Psi$,
while $S$ is the survival probability that takes into account the 
$J/\Psi$'s from the decay of the higher resonances. This calculation was done 
with the cross sections denoted (2) in Tab.~\ref{tab} and the data 
from Ref.~\cite{leitch}. 
At $x_F<0$, we find reasonable
agreement between data and calculation. At $x_F=0$ we have 
$l_c=20.6\cdot 0.3$ fm $>2\cdot R_{Be}$ and thus eq.~(\ref{expand}) is
not longer valid.

\section{Summary}
\label{sum}
\begin{itemize}
\item
The pQCD contribution into $\sigma(J/\Psi -N)$ is too small at
SPS-energies to explain $\sigma(\gamma A \rightarrow J/\Psi)$, and
increases rapidly with energy -- this could be important for RHIC and LHC.
\item
Motivated by the pQCD cross section,
the nonperturbative model for $\sigma(\bar q q)= c\cdot b^2$ predicts four different
types of charmonium nucleon
cross sections.
This reflects the fundamental property of QCD, where interactions
depend strongly on the scale.
\item
QCD predicts different cross sections for the $\chi_{lm}$-mesons,
because high energy interactions are dominated by the transverse size:
$\sigma(\chi_{11} -N)\approx 2\cdot \sigma(\chi_{10} -N)$.
\item
The production of small $\bar c c$ pairs and their
expansion are relevant for the $\Psi'$:$J/\Psi$-ratio,
which is weakly dependend on $A$ at larger Lorentz-factors.
\item
The assumption that the production of $c\bar c$-pairs is hard and not soft
could be tested by measuring the $A$-dependence of the total charm cross section.
This could also provide insight in nuclear shadowing and parton energy 
loss at large $x_F$.

\end{itemize}

\end{document}